# Optimizing the Efficiency of Accelerated Reliability Testing for the Internet Router Motherboard

Hanxiao Zhang, Department of Industrial Engineering, Tsinghua University, Beijing, China

Shouzhou Liu, Department of Industrial Engineering, Tsinghua University, Beijing, China

Yan-Fu Li, PhD, Department of Industrial Engineering, Tsinghua University, Beijing, China

*SUMMARY & CONCLUSIONS*

With the rapid development of internet Router, the complexity of its mainboard has been growing dramatically. The high reliability requirement renders the number of testing cases increasing exponentially, which becomes the bottleneck that prevents further elevation of the production efficiency. In this work, we develop a novel optimization method of two major steps to largely reduce the testing time and increase the testing efficiency. In the first step, it optimizes the selection of test cases given the required amount of testing time reduction while ensuring the coverage of failures. In the second step, selected test cases are optimally scheduled to maximize the efficiency of mainboard testing. A numerical experiment is investigated to illustrate the effectiveness of the proposed methods. The results show that the optimal subset of the test cases can be selected satisfying the 10% testing time reduction requirement, and the effectiveness index of the scheduled sequence can be improved by more than 75% with test case sequencing. Moreover, our method can self-adjust to the new failure data, which realizes the automatic configuration of board test cases.

## 1 INTRODUCTION

The internet router is a critical device of the internet infrastructure that forwards data packets between the computer networks. It is often programmed to filter packets, translate addresses, make routing decisions, broker quality of service reservations, etc. The processing speed of the router is one of the major constraints on internet speed and its reliability directly affects the quality of network service. In addition, with the rapid development and proliferation of the internet router, its main component: the motherboard, has become increasingly complex. Evidently, reliability testing of the motherboard is being ever more necessary to ensure its quality and thus to protect the reputation of the manufacturer.

The high-reliability requirement renders the number of test cases in reliability testing increasing dramatically and under the current testing scheme, all test cases must be executed at each testing period, which significantly prolongs the testing process and thus creates a bottleneck that prevents further elevation of the motherboard production. On the other hand, according to our investigation, there is a large number of test cases which have not exposed any fault in history. The observations and evidence above indicate the great potential for improvements and therefore motivate our research.

Accelerated Life Testing (ALT) [1, 2] is one of the practical and principal methods of electronic products used by manufacturers to estimate the reliability of their products. Statistical methodology and applications to estimate the failure time have been researched extensively. With the characteristic of shortening the lifetime of the products or accelerating the degradation of their performance, manufacturers can generate a fault through ALT finding the causes of the fault and correcting them before producing. On the other hand, the weakest products can be detected with ALT to reduce the rejection rate.

To guarantee that the reliability of the products can be estimated accurately and quickly, the design of a plan is critical for ALT. ALT plan consists of the chosen test conditions, the number of specimens at each test conditions and the prechosen censoring time of each specimen and so on [3, 4]. Some ALT plans consider Type II failure censoring times and periodic inspection for failure [5]. Step-stress tests can be used to yield failures quickly, which obtain the information in a short time [6]. The size of the test specimens can be determined to get more accurate results in a given time [7]. In this work, however, we focus on reducing the testing time by decreasing the number of test cases, where the ALT conditions are fixed. The selection and sequencing of the test cases are operated to meet the time reduction. These operations do not affect the lifetime of products while influencing the efficiency and effectiveness of product quality inspection.

The test case prioritization techniques: test case selection and test case sequencing, have been widely researched in the software regression test [8, 9, 11], which schedule test cases in an order to increase their effectiveness at meeting better performance goal, such as code coverage and rate of fault detection. In this paper, we draw on these principles and techniques, and modify them to adapt to our problem.

The contributions of this work are:
1. We model the problem of reducing the ALT testing time and define the notations in this model.
2. We formulate this problem into a two-step method: test case selection and test case sequencing.
3. We propose the exact algorithms to solve Test Case Selection and Test Case Sequencing respectively.

The rest of this paper is organized as follows. Section 2 introduces the description of this problem. Section 3 and 4 illustrate the models and the exact algorithms for Test Case Selection and Test Case Sequencing. Numerical experiment and results are presented in section 5. In Section 6, we conclude this work and discuss the advantages and drawback of the proposed model and methods.

## 2 THE STATEMENT OF THE PROBLEM

### 2.1 Notation

Table 1 shows the notations of the parameters we use in the following model.

*Table 1 − Denotations of parameters in our problem*

| | |
|---|---|
| $TC_{ij}$ | test case $i$ in the period $j \in J$ |
| $x_{ij}$ | decision of the number of the test case $TC_{ij}$ |
| $J$ | all periods in the ALT |
| $I_j$ | all test cases in each period $j \in J$ |
| $n_j$ | the number of test cases in each period $j \in J$ |
| $I_j^e$ | all effective test cases in each period $j \in J$ |
| $B_j, \Gamma_j$ $\Delta_j = J - I_j^e - B_j - \Gamma_j$ | three given testing sets of test cases in each period $j \in J$ |
| $p_1, p_2, p_3, p_4$ | the priority of the set $I_j^e$ and $B_j, \Gamma_j, \Delta_j$ in selection procedure individually |
| $J_{HTHV}, J_{HTLV}$ $J_{LTHV}, J_{LTLV}$ | the set of periods with the same accelerated conditions |
| $T_j$ | the expected time limit for each period $j \in J$ |
| $w_{ij}$ | the number of failures in history exposed by the test case $TC_{ij}$ |
| $T_{ij}$ | the stop time of the test case $TC_{ij}$ in the original scheduling of the testing procedure |
| $r_{ij}$ | the running time of the test case $TC_{ij}$ |
| $z_{0j}$ | the start time of each period $j$ |
| $P_j$ | precedence relation among test cases in period $j \in J$ $P_j = \{(i,k) \mid TC_{ij} < TC_{kj}\}$ |

### 2.2 Problem Description

The type of accelerated reliability test considered in this study is the temperature and voltage cycle test. It is to determine the ability of the motherboard to withstand the mechanical stresses induced by the temperature and voltage respectively alternating between two extremes. The weakest members were exposed and the data were collected for analysis.

In our testing scheme, each motherboard undergoes few cycles to expose the potential faults (Fig. 1). Each cycle consists of four different testing periods. During each period, a list of test cases is executed sequentially. A test case is a small program that can diagnose if certain functionalities of the motherboard work correctly. Thus, the failures of the motherboard should be exposed by certain test case(s). All historical failures are recorded, including the time of failure and the corresponding test cases. An effective test case is one test case that has exposed at least one fault in history.

*Fig 1 − An example for the testing conditions and the failure*

| | Cycle 1 | | | | Cycle 2 | | | |
|---|---|---|---|---|---|---|---|---|
| | Period 1 | Period 2 | Period 3 | Period 4 | Period 5 | Period 6 | Period 7 | Period 8 |
| Temperature | LT | HT | LT | HT | LT | HT | LT | HT |
| voltage | LV | LV | HV | HV | LV | LV | HV | HV |
| Test Case 1 | 0 | 5 | 0 | 5 | 0 | 5 | 0 | 5 |
| Test Case 2 | 1 | 0 | 1 | 0 | 1 | 0 | 1 | 0 |
| Test Case 3 | 0 | 0 | 0 | 0 | 0 | 0 | 0 | 0 |

*number of each test case in the cycle of ALT*

Due to the sparsity of the failure number $w_{ij}$ in the ALT, we can select a subset of test cases $I_j$ to execute with the given time limit $T_j$ and include all effective test cases to guarantee the testing efficiency. The lifetime of a product is variable with the different testing condition. Therefore, the subset of the test case is sequenced as close to the originally scheduled time as possible to increase the rate of fault detection and maintain the testing effectiveness.

In this work, we aim to develop a novel optimization method of two major steps: test case selection and test case sequencing, to essentially improve the testing efficiency while maintaining the testing effectiveness.

## 3 TEST CASE SELECTION

In the first step, the selection optimization is formulated as a modified assignment problem, solved by a linear integer programming technique. The selection of the test cases $TC_{ij}$ is coded as indicative decision variables $x_{ij}$.

$$x_{ij} = \begin{cases} nonnegative\ integer, & TC_{ij}\ is\ chosen\ in\ period\ j \\ 0, & otherwise \end{cases}$$

The objective function contains two additive parts: the first part is the sum of the number of selected test cases each weighted by the number of failures exposed in history, and another part is the sum of test cases selected in each given testing set weighted with its priority factor.

$$\max \{ \sum_{j \in J} \sum_{i \in I_j} w_{ij} x_{ij} + ( p_1 \sum_{j \in J} \sum_{i \in I_j^e} x_{ij} + p_2 \sum_{j \in J} \sum_{i \in \beta_j \setminus I_j^e} x_{ij}$$

$$+p_3 \sum_{j \in J} \sum_{i \in I_j \setminus I_j^e} x_{ij} + p_4 \sum_{j \in J} \sum_{i \in A_j} x_{ij} \,) \} \tag{1}$$

The model searches for the optimal subset of the test cases to maximize the objective function under a time limit constraint. The time limit, shown as follows, is predetermined to achieve the goal of overall testing time reduction.

$$\sum_{i \in I_j} r_{ij} x_{ij} \leq T_j \quad \forall j \in J \tag{2}$$

In addition, there are four hierarchical constraints. The present one can be included in the optimization model if there are still empty time slots to be filled after all the previous constraints are satisfied. These constraints are presented as follows:

*Constraint 1.*

The test case subset for each period must include all effective test cases which have exposed any faults in history. This is to ensure that all historical failures are covered.

$$x_{ij} \geq 1, x_{ij} \in Z^+ \quad \forall j \in J, i \in I_j^e \tag{3}$$

*Constraint 2.*

All test cases must be executed under four different periods (not necessarily in the same cycle) to ensure that each test case has experienced the four environmental conditions at least once.

$$\sum_{j \in J_k} x_{ij} \geq 1 \quad \forall J_k \in \{J_{HTHV}, J_{HTLV}, J_{LTHV}, J_{LTLV}\} \tag{4}$$

*Constraint 3.*

The effective test cases mentioned in the first constraint are allowed to appear twice in the subset for each period, in order to improve the possibility of failure detection.

$$x_{ij} \leq 2, x_{ij} \in Z^+ \quad \forall j \in J, i \in I_j^e \tag{5}$$

*Constraint 4.*

Other unselected test cases in each period can be assigned into the subset according to a predefined priority.

$$x_{ij} \in \{0,1\} \quad \forall j \in J, i \in I \setminus I_j^e \tag{6}$$

The second part of the objective function in (1) guarantees the selection priority with different priority factors $p_i$. This integer programming problem (1) - (6) can be solved with any normal integer optimization technique.

## 4 TEST CASE SEQUENCING

After the first step, we obtain an optimal subset of test cases for each testing period. In the second step, the test case sequencing is determined such that it optimizes a specific goal and simultaneously ensures that the precedence relations between certain test cases are retained.

To address our research problem, we need a measure with which can assess and compare the effectiveness of different test case sequences. Cost-cognizant weighted Average Percentage of Faults Detected measure ($APFD_C$) used in test case prioritization for software fault detection [8, 9] can be used for reference. According to the failure physics of the motherboard, the same type of failure is more likely to occur near the same point at the timeline starting from the beginning of the cycle testing. Thus, the objective $P^j$ is to minimize the total deviation between the historical failure times and the scheduled times of the effective test cases. Moreover, the failure number of each test case based on historical records is also considered as the coefficient of the deviation, which can improve the effectiveness of the decision. Lower $P^j$ for each period $j$ in our problem means better performance.

$$P^j = \frac{\sum_{i=1}^{n_j} w_{ij} |t_{ij} - T_{ij}|}{\sum_{i=1}^{n_j} r_{ij} \cdot \sum_{i=1}^{n_j} w_{ij}}$$

The model for each period $j$ is constructed as follows (7) - (14).

$$min \quad \frac{\sum_{i=1}^{n_j} w_{ij} |t_{ij} - T_{ij}|}{\sum_{i=1}^{n_j} r_{ij} \cdot \sum_{i=1}^{n_j} w_{ij}} \tag{7}$$

$$s.t. \quad x_{ik}^j + x_{ki}^j = 1 \quad \forall i, k \in I_j \tag{8}$$

$$x_{il}^j \geq x_{ik}^j + x_{kl}^j - 1 \quad \forall i, l, k \in I_j \tag{9}$$

$$x_{ik}^j = 1 \ \forall (i,k) \in P_j \tag{10}$$

$$t_{kj} - t_{ij} + M(1 - x_{ik}^j) \geq r_{kj} \quad \forall i, k \in I_j \tag{11}$$

$$t_{ij} \leq \sum_{i=1}^{n_j} r_{ij} + z_{0j} \quad \forall i \in I_j \tag{12}$$

$$t_{ij} \in Z_+ \quad \forall i \in I_j \tag{13}$$

$$x_{ik}^j \in B \quad \forall i, k \in I_j \tag{14}$$

The time when the test case $TC_{ij}$ is finished is denoted as $t_{ij}$, which is the decision variable in optimization. In (7) the objective aims at decreasing the difference between the scheduled time $t_{ij}$ and the historical failure time $T_{ij}$ of the effective test case. The incidence variable $x_{ij}$ denotes whether the test case $TC_{ij}$ is scheduled before the test case $TC_{kj}$ in period $j$.

$$x_{ik}^j = \begin{cases} 1, & \text{if } TC_{ij} \text{ is scheduled before } TC_{kj} \\ 0, & \text{otherwise} \end{cases}$$

Constraint (8) means the order between any two test cases is unidirectional. Constraint (9) guarantees the transitivity, i.e. if $x_{ik}^j \geq 1$ and $x_{kl}^j \geq 1$, then $x_{il}^j \geq 1$. Constraint (10) is also known as precedence constraint [10] in scheduling problem and it serves to limit the order of certain test cases in a sequence. Precedence constraint defines a partial ordering between the test cases: $TC_{ij} < TC_{kj}$, means that test case $TC_{kj}$ cannot start before the completion of test case $TC_{ij}$. In constraints (8) – (10) and (14), a sequence of the selected test cases is constructed. After a sequence $\{x_{ik}^j\}$ is determined, the scheduled time of each test case $t_{ij}$ is calculated in constraints (11) – (13). If $x_{ik}^j = 1$, which means the test case $TC_{ij}$ is scheduled before $TC_{kj}$, the constraint can be simplified to $t_{kj} \geq t_{ij} + r_{kj}$. Otherwise, the constraint is always satisfied and has no restriction on $t_{ij}$ and $t_{kj}$. The constraint (12) makes sure that the sequence is continuous and the testing process is never idle. To make the feasible region more compact, the value of $M$ is set as $\sum_{i=1}^{n_j} r_{ij} + z_{0j}$ in each period.

## 5 COMPUTATIONAL EXPERIMENT

To analyze the method proposed in this paper, we use a randomly generated instance. The studied ALT has two cycles and eight periods, and the accelerated conditions are the same as the example shown in Fig.1: LTLV, HTLV, LTHV, HTHV,

LTLV, HTLV, LTHV, HTHV. There are 10 test cases during ALT. The total time is 160 mins (20 mins for each period) and it is required to be reduced by 10%. The time limit $T_j$ equals to 18 mins and the start time of each period $z_{0j}$ is $18 * (j - 1)$ minute. Table 2.1 - 2.3 presents the data used for the experiment: failure numbers $w_{ij}$, test case running time $r_{ij}$, and originally scheduled time $T_{ij}$. The precedence constraints are $P_j = \{(TC3, TC8), (TC1, TCk) \ \forall \ k \in I_j\} \ \forall j \in J$. The given testing sets are $B_j = \{8\}, I_j = \{9, 10\}$ for all $j \in J$ and the priority factors $p_1, p_2, p_3, p_4$ are set as $4, 3, 2, 1$ respectively.

## 5.1 Results

The test case selection results solved by linear integer programming are presented in Table 3. We show that the time of each period is reduced by 10% with test case selection method. And the covering constraints of four accelerated conditions are also satisfied to ensure the testing effectiveness. Table 4 presents the test case sequencing results for each period.

## 5.2 Comparisons

To illustrate the effectiveness of the proposed method in the test case sequencing, we compare the results in Table 4 with the effectiveness index (7) of the sequence without optimization. The sequence without optimization uses the subset of the test cases in Table 3 and the repetitive test cases are scheduled at the end of the sequence in order. Lower objective $P^j$ means better performance. Table 5 shows that, the results obtained from test case sequencing are greater than the sequence without optimization by more than 75%, where GAP is set as (ObjVal without optimization − ObjVal with optimization) / ObjVal without optimization.

*Table 2.1 − Data of the failure number $w_{ij}$ for each test case*

| Test Case | Period1 | Period2 | Period3 | Period4 | Period5 | Period6 | Period7 | Period8 |
|-----------|---------|---------|---------|---------|---------|---------|---------|---------|
| TC1 | 20 | 20 | 15 | 14 | 5 | 3 | 1 | 2 |
| TC2 | 0 | 0 | 0 | 0 | 0 | 0 | 0 | 0 |
| TC3 | 0 | 0 | 1 | 0 | 2 | 0 | 1 | 0 |
| TC4 | 0 | 0 | 0 | 0 | 0 | 0 | 0 | 0 |
| TC5 | 1 | 0 | 0 | 0 | 1 | 0 | 0 | 0 |
| TC6 | 1 | 0 | 0 | 0 | 0 | 0 | 0 | 0 |
| TC7 | 0 | 0 | 0 | 0 | 0 | 0 | 0 | 0 |
| TC8 | 1 | 0 | 0 | 0 | 0 | 0 | 0 | 1 |
| TC9 | 0 | 0 | 0 | 0 | 0 | 0 | 0 | 0 |
| TC10 | 2 | 0 | 1 | 0 | 2 | 0 | 0 | 1 |

*Table 2.2 − Data of the running time $r_{ij}$ (seconds) for each test case*

| Test Case | Period1 | Period2 | Period3 | Period4 | Period5 | Period6 | Period7 | Period8 |
|-----------|---------|---------|---------|---------|---------|---------|---------|---------|
| TC1 | 200 | 190 | 200 | 190 | 200 | 190 | 200 | 190 |
| TC2 | 25 | 20 | 25 | 20 | 25 | 20 | 25 | 20 |
| TC3 | 100 | 150 | 200 | 210 | 100 | 150 | 200 | 210 |
| TC4 | 55 | 65 | 65 | 65 | 65 | 65 | 55 | 65 |
| TC5 | 70 | 70 | 60 | 70 | 70 | 70 | 70 | 60 |
| TC6 | 150 | 140 | 150 | 150 | 145 | 140 | 170 | 150 |
| TC7 | 125 | 120 | 120 | 110 | 115 | 120 | 115 | 110 |
| TC8 | 10 | 10 | 10 | 10 | 10 | 10 | 10 | 10 |
| TC9 | 60 | 40 | 45 | 50 | 60 | 40 | 45 | 50 |
| TC10 | 400 | 350 | 300 | 320 | 400 | 350 | 300 | 300 |

*Table 2.3 − Data of the original time $T_{ij}$ (seconds) for each test case*

| Test Case | Period1 | Period2 | Period3 | Period4 | Period5 | Period6 | Period7 | Period8 |
|-----------|---------|---------|---------|---------|---------|---------|---------|---------|
| TC1 | 200 | 1390 | 2600 | 3790 | 5000 | 6190 | 7400 | 8590 |
| TC2 | 225 | 1410 | 2625 | 3810 | 5025 | 6210 | 7425 | 8610 |
| TC3 | 325 | 1560 | 2825 | 4020 | 5125 | 6360 | 7625 | 8820 |
| TC4 | 380 | 1625 | 2890 | 4085 | 5190 | 6425 | 7680 | 8885 |
| TC5 | 450 | 1695 | 2950 | 4155 | 5260 | 6495 | 7750 | 8945 |
| TC6 | 600 | 1835 | 3100 | 4305 | 5405 | 6635 | 7920 | 9095 |
| TC7 | 725 | 1955 | 3220 | 4415 | 5520 | 6755 | 8035 | 9205 |
| TC8 | 735 | 1965 | 3230 | 4425 | 5530 | 6765 | 8045 | 9215 |
| TC9 | 795 | 2005 | 3275 | 4475 | 5590 | 6805 | 8090 | 9265 |
| TC10 | 1195 | 2355 | 3575 | 4795 | 5990 | 7155 | 8390 | 9565 |

*Table 3 − Test case selection results*

| Test Case | Period1 | Period2 | Period3 | Period4 | Period5 | Period6 | Period7 | Period8 |
|---|---|---|---|---|---|---|---|---|
| TC1 | 2 | 2 | 2 | 2 | 1 | 2 | 1 | 2 |
| TC2 | 1 | 1 | 1 | 1 | 0 | 1 | 1 | 1 |
| TC3 | 0 | 1 | 1 | 1 | 1 | 0 | 2 | 0 |
| TC4 | 0 | 1 | 1 | 1 | 1 | 1 | 1 | 0 |
| TC5 | 1 | 1 | 1 | 1 | 2 | 1 | 0 | 1 |
| TC6 | 1 | 1 | 0 | 1 | 0 | 0 | 1 | 0 |
| TC7 | 0 | 1 | 0 | 1 | 1 | 1 | 1 | 0 |
| TC8 | 2 | 1 | 1 | 1 | 0 | 1 | 1 | 2 |
| TC9 | 0 | 0 | 0 | 1 | 1 | 1 | 1 | 0 |
| TC10 | 1 | 0 | 1 | 0 | 1 | 1 | 0 | 2 |
| Total time | 1065 | 955 | 1060 | 1065 | 1080 | 1055 | 1020 | 1080 |

*Table 4 − Test case sequencing results ("TC1-" denotes the same test case as TC1)*

| order | Period1 | Period2 | Period3 | Period4 | Period5 | Period6 | Period7 | Period8 |
|---|---|---|---|---|---|---|---|---|
| 1 | TC1 | TC1 | TC1 | TC1 | TC1 | TC1 | TC1 | TC1 |
| 2 | TC1- | TC1- | TC1- | TC4 | TC7 | TC7 | TC2 | TC2 |
| 3 | TC5 | TC6 | TC4 | TC3 | TC9 | TC4 | TC7 | TC10- |
| 4 | TC6 | TC3 | TC3 | TC7 | TC5- | TC10 | TC4 | TC8- |
| 5 | TC8- | TC4 | TC5 | TC8 | TC4 | TC1- | TC9 | TC5 |
| 6 | TC2 | TC5 | TC8 | TC6 | TC5 | TC9 | TC3- | TC1- |
| 7 | TC8 | TC8 | TC2 | TC9 | TC3 | TC5 | TC6 | TC10 |
| 8 | TC10 | TC2 | TC10 | TC1- | TC10 | TC8 | TC3 | TC8 |
| 9 | -- | TC7 | -- | TC5 | -- | TC2 | TC8 | -- |
| 10 | -- | -- | -- | TC2 | -- | -- | -- | -- |
| ObjVal | 1.3897 | 12.5654 | 21.9478 | 33.8028 | 38.7963 | 56.8720 | 41.4216 | 73.8426 |

*Table 5 − The comparison results: scheduled sequence and original sequence*

| ObjVal (%) | Period1 | Period2 | Period3 | Period4 | Period5 | Period6 | Period7 | Period8 |
|---|---|---|---|---|---|---|---|---|
| without optimization | 5.9415 | 54.9020 | 95.5198 | 144.0000 | 162.0297 | 242.7746 | 307.3171 | 458.6207 |
| with optimization | 1.3897 | 12.5654 | 21.9478 | 33.8028 | 38.7963 | 56.8720 | 41.4216 | 73.8426 |
| Gap (%) | 76.6108 | 77.1129 | 77.0227 | 76.5258 | 76.0561 | 76.5741 | 86.5216 | 83.8990 |

## 6  CONCLUSIONS

In conclusion, our two-step method produces the optimally selected and optimally ordered subsets of test cases, which can maximize the effectiveness of the cycle testing and reduce the testing time to the desired level. Our algorithm for test case sequencing is used for the small size of test cases and can obtain the exact solution. In an industrial process, we can develop evolutionary algorithms to optimize the test case sequence [11, 12, 13, 14].

Our method has been successfully applied to the router motherboard production of a major Chinese telecommunication manufacturer, satisfying the 20% testing time reduction requirement while scheduling the effective test cases around their historical failure times. Moreover, our methodology can self-adjust to the new failure data and eventually realizes the automation of the optimal selection and sequencing of the motherboard reliability testing. The promising results indicate its applicability to other similar reliability testing processes.

*BIOGRAPHIES*

Hanxiao Zhang
Department of Industrial Engineering
Tsinghua University
30 Shuangqing Road
Beijing, China

e-mail: zhx17@mails.tshinghua.edu.cn

Hanxiao Zhang received a B.S. degree in Applied mathematics from Wuhan University, China, in 2017. She is currently a Ph.D. candidate in the Department of Industrial Engineering, Tsinghua University. Her research interests include Redundancy Allocation Problem, dynamic programming and optimization methods.

Shouzhou Liu
Department of Industrial Engineering
Tsinghua University
30 Shuangqing Road
Beijing, China

e-mail: lsz16@mails.tsinghua.edu.cn

Shouzhou Liu received a M.S. degree in Industrial Engineering from Tsinghua University, China, in 2019. His research interests include smart grid, cyber security and game theory.

Yan-Fu Li, PhD
Department of Industrial Engineering
Tsinghua University
30 Shuangqing Road
Beijing, China

e-mail: liyanfu@tsinghua.edu.cn

Yan-Fu Li is currently a professor at the Department of Industrial Engineering (IE), Tsinghua University. He is the director of the Reliability & Risk Management Laboratory at Institute of Quality and Reliability in Tsinghua University. He obtained his B.Eng. degree in software engineering from Wuhan University in 2005 and a Ph.D. in industrial engineering from the National University of Singapore in 2010. He was a faculty member at the Laboratory of Industrial Engineering at CentraleSupélec, France, from 2011 to 2016. His current research areas include RAMS (reliability, availability, maintainability, safety and security) assessment and optimization with the applications onto energy systems, transportation systems, computing systems, etc. He is the Principal Investigator on several government projects including one key project funded by National Natural Science Foundation of China, one project in National Key R&D Program of China, and the projects supported by EU and French funding bodies. He is also experienced in industrial research: partners include EDF, ALSTOM, China Southern Grid, etc. Dr. Li has published more than 90 research papers, including more than 40 peer-reviewed international journal papers. Dr. Li is currently an associate editor of IEEE Transactions on Reliability, a senior member of IEEE and a member of INFORMS. He is a member of the Executive Committee of the Reliability Chapter of Chinese Operations Research Society; Executive Committee of Industrial Engineering Chapter of Chinese Society of Optimization, Overall Planning and Economic Mathematics; Committee of Uncertainty Chapter of Chinese Artificial Intelligence Society.